\documentclass[a4paper,12pt]{article}

\usepackage[utf8]{inputenc}
\usepackage{comment}
\usepackage[margin=2.5cm]{geometry}
\usepackage{graphicx}
\usepackage{caption}
\usepackage{subfig}
\usepackage{hyperref}
\usepackage{amsmath}
\usepackage[dvipsnames]{xcolor}
\usepackage{authblk}

\usepackage{multirow}

\title{Quasinormal modes of rotating black holes in shift-symmetric Einstein-scalar-Gauss-Bonnet theory}

\author[1,2]{Fech Scen Khoo\thanks{\href{mailto:fkhoo@ucm.es}{fkhoo@ucm.es}}}
\author[1]{Jose Luis Bl\'azquez-Salcedo\thanks{\href{mailto:jlblaz01@ucm.es}{jlblaz01@ucm.es}}} 
\author[2]{Burkhard Kleihaus\thanks{\href{mailto: b.kleihaus@uni-oldenburg.de}{b.kleihaus@uni-oldenburg.de}}} 
\author[2]{Jutta Kunz\thanks{\href{mailto:jutta.kunz@uni-oldenburg.de}{jutta.kunz@uni-oldenburg.de}}}  
\affil[1]{Departamento de F\'isica Te\'orica and IPARCOS, Facultad de Ciencias F\'isicas, Universidad Complutense de Madrid, 28040 Madrid, Spain}
\affil[2]{Institut f\"ur  Physik, Universit\"at Oldenburg, Postfach 2503,
D-26111 Oldenburg, Germany}

\date{\today}

\begin{document}

\maketitle

\begin{abstract}
We employ a recently developed spectral method to obtain the spectrum of quasinormal modes 
of rapidly rotating black holes in alternative theories of gravity and apply it to the black holes of shift-symmetric Einstein-scalar-Gauss-Bonnet theory.
In this theory the quasinormal modes were recently obtained by employing perturbation theory in quadratic order in the Gauss-Bonnet coupling constant.
Here we present the full non-perturbative results for the spectrum within the domain of existence of rotating black holes and compare with the perturbative results.
We also compare with the quasinormal mode spectrum of rapidly rotating Einstein-dilaton-Gauss-Bonnet black holes.
\end{abstract}

\section{Introduction}

The detection of gravitational waves from the inspiral, merger and ringdown of compact objects provides us with an excellent new tool to study the strong gravity regime \cite{LIGOScientific:2016aoc,Cahillane:2022pqm}.
The numerical relativity predictions obtained within General Relativity (GR) match well the LIGO/Virgo/KAGRA observations, which so far support the \textit{Kerr paradigm}, i.e., the expectation that rotating black holes in the Universe are well described in terms of the (exterior) Kerr black hole solution.
The next generation of gravitational wave detectors like the Einstein Telescope, the Cosmic Explorer or LISA will have larger signal-to-noise ratios together with larger event numbers  (see e.g., \cite{Punturo:2010zz,Dwyer:2015,Colpi:2024xhw}).
Thus the Kerr paradigm will be subject to scrutiny.

There are various reasons to expect that GR will be superseded by a more complete theory of gravity.
Thus in recent years numerous alternative theories of gravity have been suggested, driven largely by cosmological considerations
and attempts to reconcile GR with quantum mechanics
\cite{CANTATA:2021ktz, Berti:2015itd}.
Of course, any viable gravitational theory should conform with all experiments and observations \cite{Will:2018bme}.
But currently the strong gravity sector is still in need of far more testing \cite{Berti:2015itd}.
Here the next generation of gravitational wave detectors should play an important role.

After the merger of two black holes the newly formed highly excited black hole undergoes ringdown where it emits damped sinusoidal gravitational waves, while it evolves towards a stationary state.
These waves are associated with the quasinormal modes  \cite{Kokkotas:1999bd,Berti:2009kk,Konoplya:2011qq} of the final black hole and are dominated by the longest-lived ones
(although nonlinearities in the black-hole ringdown could also be important \cite{Mitman:2022qdl,Cheung:2022rbm}).
Like the black holes themselves, the quasinormal modes depend on the gravity theory.
Therefore accurate ringdown measurements have the potential to put stringent bounds on alternative gravity theories 
(see e.g. \cite{Maenaut:2024oci, Chung:2025wbg}).

The calculation of the quasinormal mode spectrum for Kerr black holes was achieved long ago \cite{Teukolsky:1973ha}.
But already the quasinormal mode spectrum of the Kerr-Newman black holes represented a challenging problem, solved only in the last decade \cite{Dias:2015wqa,Dias:2021yju,Dias:2022oqm}.
For alternative theories of gravity, however, the techniques to obtain the spectra of rapidly rotating black holes remained lacking until recently, and only some perturbative results were known for slowly rotating black holes,
with notable exceptions in \cite{Cano:2023tmv,Cano:2023jbk,Cano:2024ezp}, where the expansion in the spin parameter was taken to rather high order, and in
\cite{Bohra:2023vls,Miguel:2023rzp}, where scalar and vector modes
were calculated for rapidly rotating black holes.

Here we report the spectrum of rapidly rotating black holes in shift-symmetric Einstein-scalar-Gauss-Bonnet (EsGB) theory.
This gravity theory belongs to the class of Horndeski gravities  \cite{Horndeski:1974wa}, which feature an additional scalar degree of freedom, are free of ghosts, and give rise to second order field equations.
In EsGB theories the scalar field $\varphi$ is coupled with coupling strength $\alpha$ and  coupling function $f(\varphi)$ to the Gauss-Bonnet invariant.
In shift-symmetric EsGB theory this coupling function is simply linear in the scalar field, $f(\varphi)=\varphi$.
The theory is thus invariant under a shift of the scalar by a constant.

The static black holes of this theory were obtained by Sotiriou and Zhou \cite{Sotiriou:2013qea,Sotiriou:2014pfa}, and the rotating black holes by Delgado et al.~\cite{Delgado:2020rev}  (see also Sullivan et al.~\cite{Sullivan:2020zpf}).
Delgado et al.~mapped carefully the domain of existence of rotating shift-symmetric EsGB black holes.
This domain must, of course, be respected in the calculation of the quasinormal modes, since beyond this domain no such black holes exist.
The quasinormal modes of rapidly rotating shift-symmetric EsGB black holes were obtained recently by Chung and Yunes by applying perturbation theory in the coupling constant to quadratic order in $\alpha$ \cite{Chung:2024ira,Chung:2024vaf}
(see also \cite{Hirano:2024pmk} for a recent study on the quasinormal modes of the static configurations).
We here present the full results, based on an exact treatment of the dependence on the coupling constant $\alpha$ both for the background solutions and the perturbations, and then compare these exact results with the perturbative results.
We note that black hole-neutron star mergers in this theory were also recently studied \cite{Corman:2024vlk}.

To obtain the quasinormal modes we employ our recently developed spectral scheme \cite{Blazquez-Salcedo:2023hwg,Khoo:2024yeh,Blazquez-Salcedo:2024oek, Blazquez-Salcedo:2024dur}.
After testing this scheme extensively for Kerr black holes \cite{Blazquez-Salcedo:2023hwg}, we applied it to the rapidly rotating black holes of Einstein-dilaton-Gauss-Bonnet (EdGB) theory, a string theory motivated alternative theory of gravity \cite{Gross:1986mw,Metsaev:1987zx}.
Its hairy black holes are well-studied \cite{Kanti:1995vq,Torii:1996yi,Guo:2008hf,Pani:2009wy,Kleihaus:2011tg,Pani:2011gy,Ayzenberg:2013wua,Ayzenberg:2014aka,Kleihaus:2014lba,Maselli:2015tta,Kleihaus:2015aje,Blazquez-Salcedo:2016enn,Cunha:2016wzk,Zhang:2017unx, Blazquez-Salcedo:2017txk,Ripley:2019irj,Pierini:2021jxd,Pierini:2022eim}, and also the emission of gravitational waves during the inspiral, merger and ringdown of binary systems was recently addressed \cite{Julie:2024fwy}.

Since EdGB theory has an exponential coupling function $f(\varphi)=\exp{(\varphi)}$, the shift-symmetric EsGB theory may be considered to be a linearization of the EdGB theory.
From that point of view one may expect similarities between the black holes of these theories and between their quasinormal modes. Indeed, the domains of existence of rotating black holes are rather similar in both theories, with the domain in shift-symmetric EsGB theory extending further in the coupling constant $\alpha$ \cite{Kleihaus:2011tg,Delgado:2020rev}. Here we also compare the quasinormal mode spectra of both theories.

\section{Rotating black holes}
\label{setup}

We now briefly present the action and the field equations of shift-symmetric EsGB theory and then recall the construction and properties of the rotating black holes in this theory, first obtained in Delgado et al.~\cite{Delgado:2020rev}.

\subsection{Action and field equations}

We here consider the shift-symmetric EsGB theory
\begin{equation}
    S(g,\varphi) = \frac{1}{16\pi} \int d^4x \sqrt{-g}
    \left(R - \frac{1}{2}\partial_{\mu}\varphi \, \partial^{\mu}\varphi 
    +  \alpha \varphi R^2_{\text{GB}}
    \right)\, ,
    \label{action}
\end{equation}
with scalar field $\varphi$,
Gauss-Bonnet coupling constant $\alpha$,
and Gauss-Bonnet invariant
\begin{equation}
    R^2_{\text{GB}} = R_{\mu\nu\rho\sigma}R^{\mu\nu\rho\sigma}
- 4 R_{\mu\nu}R^{\mu\nu} + R^2 \,.
\end{equation}

Variation of the action (\ref{action}) leads to the generalized set of Einstein equations
\begin{equation}
    G_{\mu\nu} 
    - \frac{1}{2}T_{\mu\nu}^{(\varphi)}
    + \frac{1}{2}T_{\mu\nu}^{(GB)} = 0 \, ,
\end{equation}
with
\begin{eqnarray}
    G_{\mu\nu} &=& R_{\mu\nu} - \frac{1}{2}g_{\mu\nu} R \, ,\nonumber \\
    T_{\mu\nu}^{(\varphi)} &=&
    \nabla_{\mu}\varphi \nabla_{\nu} \varphi
    - \frac{1}{2} g_{\mu\nu}(\nabla\varphi)^2 \,,
    \nonumber \\
     T_{\mu\nu}^{(GB)}
     &=&   2 \alpha P_{\mu\sigma\nu\rho} \nabla^{\rho}\nabla^{\sigma} \varphi \,,
     \nonumber      \\
     P_{\mu\nu\rho\sigma} &=& R_{\mu\nu\rho\sigma}
     + 2 g_{\mu[\sigma} R_{\rho]\nu} 
     + 2  g_{\nu[\rho} R_{\sigma]\mu}
     + R g_{\mu[\rho} g_{\sigma]\nu} \,,
\end{eqnarray}
and to the scalar field equation
\begin{equation}
\nabla_\mu \nabla^\mu \varphi + \alpha R^2_{\rm GB}  =0 \ .
\label{scalareq}
\end{equation}

\subsection{Stationary rapidly rotating background black holes}

In order to construct the axially symmetric stationary background black hole solutions we employ the line element parametrization \cite{Kleihaus:2000kg}
\begin{eqnarray}
\label{metric}
ds^2=- f dt^2 +  \frac{m}{f} \left( d r^2+ r^2d\theta^2 \right) 
 +  \frac{\ell}{f} r^2\sin^2\theta (d\phi-\frac{\omega}{r} dt)^2 ,
\end{eqnarray}
with ``quasi-isotropic" spherical coordinates. 
The metric functions $f$, $m$, $\ell$ and $\omega$ depend only on the coordinates $r$ and $\theta$, and so does the scalar field $\varphi$,
\begin{eqnarray}
\label{scalar}
\varphi=\varphi(r,\theta).
\end{eqnarray}

The coupled set of field equations is solved numerically subject to the boundary conditions, that ensure asymptotically flat solutions which are regular everywhere on and outside the event horizon.
Consequently, we require at radial infinity
\begin{eqnarray}
f|_{r=\infty}= m|_{r=\infty}= \ell|_{r=\infty}=1 \ , \ \ \
\omega|_{r=\infty}=\phi|_{r=\infty}= 0
\ , \label{bc1a} 
\end{eqnarray}
at the horizon with horizon radius $r_{\rm H}$
\begin{eqnarray}
 \label{bc-horizon} 
f|_{r=r_{\rm H}}=
m|_{r=r_{\rm H}}=
\ell|_{r=r_{\rm H}}=0
\ , \ \ \ \omega|_{r=r_{\rm H}}=\omega_{\rm H}, \ \ \ \partial_r \phi|_{r=r_{\rm H}}=0 \,,
\end{eqnarray}
where $\omega_{\rm H}$ is a constant, 
and on the symmetry axis,
$i.e.$ at $\theta=0,\pi$,
\begin{eqnarray}
\label{bc-axis}
& &\partial_\theta f|_{\theta={0,\pi}} =
   \partial_\theta m|_{\theta={0,\pi}} =
   \partial_\theta \ell|_{\theta={0,\pi}} =
   \partial_\theta \omega|_{\theta={0,\pi}} = 
   \partial_\theta \phi|_{\theta={0,\pi}} =
   0 \, . 
\end{eqnarray}
Since the solutions possess a $Z_2$ symmetry $w.r.t.$ 
reflection on the equatorial plane, calculations can be restricted to the angular range $0\leq \theta \leq \pi/2$ by imposing on the equatorial plane the boundary conditions
\begin{eqnarray}
\label{bc-equator}
\partial_\theta f|_{\theta=\pi/2} =
 \partial_\theta m|_{\theta=\pi/2} =
  \partial_\theta \ell|_{\theta=\pi/2} =
   \partial_\theta \omega|_{\theta=\pi/2} =
	\partial_\theta \phi|_{\theta=\pi/2} =0 \, .
\end{eqnarray}

The global charges of the solutions, mass $M$ and angular momentum $J$ can be read from the asymptotic behavior of the metric for large $r$, while the asymptotic dependence of the scalar field yields the scalar charge $Q$,
\begin{eqnarray}
   g_{tt}(r,\theta) \approx 1 - \frac{2M}{r} + ...  \ \, ,\   
   g_{t\phi}(r,\theta) \approx -\frac{2J}{r}\sin^2{\theta}  + ...\ \, ,\
   \varphi(r,\theta) \approx \frac{Q}{r} + ...  \, .
\end{eqnarray}
The parameter $\omega_{\rm H}$ determines the event horizon angular velocity $\Omega_{\rm H}= \omega_{\rm H}/r_{\rm H}$ for fixed horizon size $r_{\rm H}$.

\begin{figure}[t]%
\begin{center}
  \includegraphics[width=10.0cm]{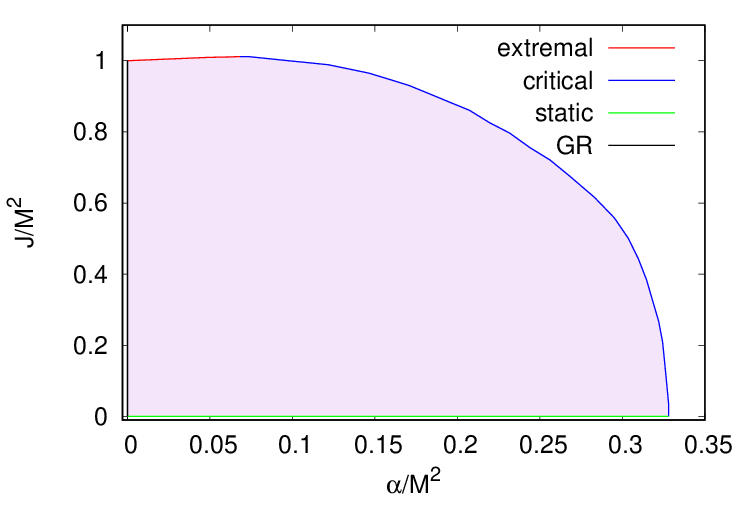}
  \caption{The domain of existence of rotating black hole solutions in shift-symmetric EsGB theory showing the dimensionless angular momentum $J/M^2$ versus the dimensionless coupling constant $\alpha/M^2$.
  The boundary consists of extremal, critical, static, and Kerr black holes. (Data extracted from \cite{Delgado:2020rev}).}
  \label{fig1}
\end{center}
\end{figure}

It is essential to know the
domain of existence of black holes in shift-symmetric EsGB theory
since beyond this domain the theory no longer supports black holes.
Perturbation theory in the coupling constant $\alpha$ yields results that are not limited per se, however, they deteriorate strongly towards this boundary and become meaningless beyond the boundary.
Therefore we demonstrate this domain of rotating background solutions in figure \ref{fig1}, which shows the dimensionless angular momentum $J/M^2$ versus the dimensionless coupling constant $\alpha/M^2$ \cite{Delgado:2020rev}.

The boundary of the domain of existence is provided by several distinct sets of solutions.
The left boundary (black, axis) corresponds to the solutions of General Relativity, the Schwarzschild black hole (origin) and the Kerr black holes (axis).
The lower boundary (green, axis) represents the set of static shift-symmetric EsGB black holes.
The red part of the upper boundary shows the extremal black holes of this theory, which can slightly exceed the Kerr bound $J/M^2=1$.
Finally, the blue (solid) curve shows the critical black holes.
These correspond to solutions, where some discriminant vanishes, and therefore represent endpoints in the domain.

\section{Metric and scalar perturbations}

We now specify the set of perturbations for the metric and the scalar field and fix the gauge.
Then we address the boundary conditions, choose new coordinates and apply a spectral decomposition, to obtain the final set of equations to be solved numerically.

\subsection{Ansatz}
\label{sec_ansatz}

We decompose the full metric into the background part $g^{(bg)}_{\mu\nu}$ and the perturbations $\delta h_{\mu\nu}(t,r,\theta,\phi)$
\begin{eqnarray}
g_{\mu\nu} &=& g^{(bg)}_{\mu\nu} + \epsilon \delta h_{\mu\nu}(t,r,\theta,\phi) 
= g^{(bg)}_{\mu\nu} + \epsilon \left(\delta h^{(A)}_{\mu\nu} + \delta h^{(P)}_{\mu\nu}\right)  \, , 
\end{eqnarray}
where $\epsilon$ is the perturbation parameter, and
the superscripts $(P)$ and $(A)$ denote the axial-led and polar-led parts, given by
\begin{equation}
 \delta h^{(A)}_{\mu\nu} = e^{i(M_z\phi-\omega t)} 
\begin{pmatrix}
0             & 0             & a_1(r,\theta) & a_2(r,\theta) \\
0             & 0             & a_3(r,\theta) & a_4(r,\theta) \\
a_1(r,\theta) & a_3(r,\theta) & 0             & 0 \\
a_2(r,\theta) & a_4(r,\theta) & 0             & 0
\end{pmatrix}   
\label{matrix_a}
\end{equation}
and
\begin{equation}
 \delta h^{(P)}_{\mu\nu} = e^{i(M_z\phi-\omega t)} 
\begin{pmatrix}
N_0(r,\theta) & H_1(r,\theta) & 0             & 0 \\
H_1(r,\theta) & L_0(r,\theta) & 0             & 0 \\
0             & 0             & T_0(r,\theta) & 0  \\
0             & 0             & 0             & S_0(r,\theta) 
\end{pmatrix}   \, .
\label{matrix_p}
\end{equation}
The remaining symmetries allow us to extract the dependence on the azimuthal angle $\phi$ and time $t$ by introducing the azimuthal number $M_z$ and the complex mode eigenvalue $\omega$.
Likewise, we decompose the scalar field into background and perturbation
\begin{eqnarray}
\varphi &=& \varphi^{(bg)} + \epsilon \delta\varphi(t,r,\theta,\phi) = \varphi^{(bg)} + \epsilon e^{i(M_z\phi-\omega t)} \Phi(r,\theta) \, .
\end{eqnarray}
The background metric and scalar field are given by (\ref{metric}) and (\ref{scalar}) respectively. A more thorough description of the background solutions can be found in \cite{Delgado:2020rev}.

Insertion of the Ansatz for the metric and the scalar field leads to the coupled set of metric and scalar equations
\begin{eqnarray}
\mathcal{G}_{\mu\nu} = \mathcal{G}_{\mu\nu}^{(bg)} + \epsilon \delta\mathcal{G}_{\mu\nu} (r,\theta) e^{i(M_z\phi-\omega t)}  =0 \, , \\
\mathcal{S} = \mathcal{S}^{(bg)} + \epsilon \delta\mathcal{S} (r,\theta) e^{i(M_z\phi-\omega t)}   =0 \, .
\end{eqnarray}

\subsection{Equations and boundary conditions}

It is convenient to 
redefine the 
set of functions
$\{a_1,a_2,a_3,a_4,N_0,H_1,L_0,T_0,S_0\}$
in (\ref{matrix_a}) and (\ref{matrix_p}) 
 to simplify the equations and to fix the gauge
 (see e.g.~\cite{Blazquez-Salcedo:2023hwg}).
Three pairs of the metric functions can be redefined using three functions, hence reducing the number of independent metric functions by three. The redefinitions are
\begin{eqnarray}
a_1(r,\theta) &=& - i M_z \frac{h_0(r,\theta)}{\sin{\theta}} \, , \\
a_2(r,\theta) &=& \sin{\theta} \, \partial_\theta h_0(r,\theta) \, ,  \\
a_3(r,\theta) &=& - i M_z \frac{h_1(r,\theta)}{\sin{\theta}} \, ,  \\
a_4(r,\theta) &=& \sin{\theta} \, \partial_\theta h_1 \, ,  \\
N_0(r,\theta) &=& \left( g^{(bg)}_{rr}(r,\theta) \right)^{-1}  N(r,\theta) \, ,  \\
L_0(r,\theta) &=& \left( g^{(bg)}_{rr}(r,\theta) \right)  L(r,\theta) \, ,  \\
T_0(r,\theta) &=& \left( g^{(bg)}_{\theta\theta}(r,\theta) \right) T(r,\theta)  \, , \\
S_0(r,\theta) &=& \left( g^{(bg)}_{\phi\phi}(r,\theta) \right) T(r,\theta) \, ,
\end{eqnarray}
in addition to the function $H_1$ that remains.
Thus we retain a set of seven independent functions which we collect in a function vector $\vec X=
[\widetilde H_1, \widetilde T, \widetilde N, \widetilde L, \widetilde h_0, \widetilde h_1, \widetilde \Phi_1]$ 
that is composed of six of the metric perturbation functions plus the scalar one,
and where we have used the following parametrization,
\begin{eqnarray}
 H_1 &=& \widetilde H_1(x,y)  \, \frac{1}{x(1-x)}   \, e^{i \hat{R}} \, ,
 \label{H1}
 \\
 T &=& \widetilde T(x,y)    \,  e^{i \hat{R}} \, , \\
 N &=& \widetilde N(x,y)  \, \frac{1}{1-x}    \, e^{i \hat{R}} \, , \\
 L &=& \widetilde L(x,y) \,  \frac{1}{x^2(1-x)}   \, e^{i \hat{R}} \, , \\
 h_0 &=& \widetilde h_0(x,y) \,  \frac{1}{1-x}   \, e^{i \hat{R}} \, , \\
 h_1 &=& \widetilde h_1(x,y) \,  \frac{1}{x(1-x)}   \, e^{i \hat{R}} \, , \\
 \Phi &=& \widetilde \Phi_1(x,y)  \, (1-x)  \,  e^{i \hat{R}} \, ,
   \label{phi_param}
\end{eqnarray}
where the function $\hat{R}$ is chosen such that the perturbations satisfy the ingoing/outgoing wave conditions that are explained below.

Furthermore, we introduce a new set of coordinates, $x$ and $y$,
\begin{eqnarray}
    x = \frac{r-r_H}{{r}+1} \, , \, \, \,   y = \cos\theta \, .
\end{eqnarray}
The radial coordinate is thus compactified, $0 \le x \le 1$, with the horizon located at $x=0$, and asymptotic infinity at $x=1$.
The angular coordinate $y$ resides in $-1 \le y \le 1$, with the symmetry axis located at $y=\pm 1$. 
In terms of the function vector $\vec X$ and the new coordinates $x$ and $y$ we then obtain a lengthy system of seven linear homogeneous partial differential equations, which we abbreviate as follows
\begin{eqnarray}
    \mathcal{D}_{\mathrm{I}}(x,y) \vec{X}(x,y) = 0, \, \, \, \, \quad   \mathrm{I} = 1,...,7 \, .
    \label{metric_eq_xy}
\end{eqnarray}

Next we turn to the boundary conditions, since we must ensure, that the perturbation functions are purely outgoing waves at infinity and purely ingoing waves at the horizon.
To this end we factor out of the perturbation functions the common factor $\exp{(i\omega R^*)}$ with eigenvalue $\omega$ and radial function $R^*$, 
requiring the proper choice of $R^*$ at the boundaries of the integration,
as follows.
At infinity, the perturbation functions are described by
\begin{eqnarray}
           T &=& e^{i\omega R^* } \left( T^{+}(\theta) + \mathcal{O}\left(\frac{1}{r}\right) \right) \, ,\\
           H_{1} &=& r e^{i\omega R^* } \left( H^{+}_{1}(\theta) + \mathcal{O}\left(\frac{1}{r}\right) \right)\, ,\\
           N &=& r e^{i\omega R^* } \left( N^{+}(\theta) + \mathcal{O}\left(\frac{1}{r}\right)  \right) \, ,\\
           L &=& r e^{i\omega R^* } \left( L^{+}(\theta) + \mathcal{O}\left(\frac{1}{r}\right)  \right) \, ,\\
           h_{0} &=&  r e^{i\omega R^* } \left( h^{+}_{0}(\theta) + \mathcal{O}\left(\frac{1}{r}\right) \right)\, ,\\
           h_{1} &=&  r e^{i\omega R^* } \left( h^{+}_{1}(\theta)  + \mathcal{O}\left(\frac{1}{r}\right) \right)\, ,\\
           \Phi_{1} &=& \frac{1}{r} e^{i\omega R^* } \left(  \Phi_{1}^{+}(\theta) + \mathcal{O}\left(\frac{1}{r}\right)   \right)\, ,
\end{eqnarray}
and at the horizon,
\begin{eqnarray}
           T &=& e^{-i(\omega-2\Omega_H) R^* } \left( T^{-}(\theta)  + 
           \mathcal{O}\left(r-r_H\right) \right)\, ,\\
           H_{1} &=& \frac{r_H}{r-r_H} e^{-i(\omega-2\Omega_H) R^* }  \left( H^{-}_{1}(\theta) + \mathcal{O}\left(r-r_H\right) \right)\, ,\\
           N &=&  e^{-i(\omega-2\Omega_H) R^* } \left( N^{-}(\theta)  + \mathcal{O}\left(r-r_H\right) \right)\, ,\\
           L &=& \frac{r_H^2}{(r-r_H)^2} e^{-i(\omega-2\Omega_H) R^* }  \left( L^-(\theta) + \mathcal{O}\left(r-r_H\right) \right)\, ,\\
           h_{0} &=& e^{-i(\omega-2\Omega_H) R^* }  \left( h^{-}_{0}(\theta) + \mathcal{O}\left(r-r_H\right) \right)\, ,\\
           h_{1} &=& \frac{r_H}{r-r_H} e^{-i(\omega-2\Omega_H) R^* }  \left( h^{-}_{1}(\theta) + \mathcal{O}\left(r-r_H\right) \right)\, ,\\
           \Phi_{1} &=& \frac{1}{r} e^{-i(\omega-2\Omega_H) R^* }  \left(  \Phi_{1}^-(\theta) + \mathcal{O}\left(r-r_H\right) \right)\, , 
\end{eqnarray}
where $\Omega_H = \frac{\sqrt{M^2-4r_H^2}}{2M(M+2r_H)}$.
Towards radial infinity we thus require $\frac{dR^*}{dr} =  1 + \frac{2M}{r} + \mathcal{O}\left(\frac{1}{r^2}\right)$, 
and towards the horizon $\frac{dR^*}{dr} = \frac{g_1}{(r-r_H)} + \mathcal{O}(1)$, 
where the $g_1$ parameter is obtained from the horizon expansion of the background functions and corresponds to the inverse surface gravity 
$\kappa_H^{-1}$
as in the case of Kerr black holes \cite{Damour:1976jd}, the perturbative shift-symmetric EsGB black holes \cite{Chung:2024ira,Chung:2024vaf} and EdGB black holes \cite{Blazquez-Salcedo:2024oek, Blazquez-Salcedo:2024dur},
i.e. $g_1=\kappa_H^{-1}$.
After introducing these expressions into the field equations, it is then straightforward to obtain the boundary conditions that the perturbation functions have to satisfy at each boundary.
These boundary conditions can then be given in operator form, e.g., they read
\begin{eqnarray}
    \mathcal{A}_{\mathrm{I}}(x,y) \vec{X}(x,y)|_{x=0} = 0, \, \, \, 
    \quad 
    \mathrm{I} = 1,...,7 \, , \\
    \mathcal{B}_{\mathrm{I}}(x,y) \vec{X}(x,y)|_{x=1} = 0, \, \, \, 
    \quad 
    \mathrm{I} = 1,...,7 \, ,
\label{bcg_inf}
\end{eqnarray}
where $\mathcal{A}_{\mathrm{I}}, \mathcal{B}_{\mathrm{I}}$ are linear operators in $(x,y)$
(see \cite{Blazquez-Salcedo:2024dur}).

\subsection{Spectral decomposition}

In order to calculate the quasinormal modes we now decompose the perturbation functions $X_I$, collected in the vector $\vec X$ in terms of suitable well-known functions.
We expand the radial part of the perturbation functions in terms of Chebyshev polynomials of the first kind, $T_{k}(x)$, and we expand the angular part of the perturbation functions in terms of Legendre polynomials of the first kind, $P_l^{M_z}(y)$ \cite{Blazquez-Salcedo:2023hwg}.
Then all seven perturbation functions can be expressed as follows
\begin{eqnarray}
     \widetilde X_{\mathrm{I}}(x,y) &=& \sum_{k=0}^{N_x-1}  \, \, \sum_{l=|M_z|}^{N_y+|M_z|-1} C_{{\mathrm{I}},k,l}  \,  T_{k}(x)  \,  P_l^{M_z}(y)   \, ,
\end{eqnarray}
for ${\mathrm{I}}=1,...,7$.
The functions  $\widetilde X_{\mathrm{I}} =[\widetilde H_1, \widetilde T, \widetilde N, \widetilde L, \widetilde h_0, \widetilde h_1, \widetilde \Phi_1]$ 
as seen from (\ref{H1})-(\ref{phi_param})
are those that remain after a re-parametrization of the perturbation functions where the appropriate ingoing
($\sim e^{-i(\omega-2\Omega_H) R^* }$)
and
outgoing 
($\sim e^{i\omega R^* }$)
behaviors have been factored out, hence ensuring that our resulting perturbation system 
has the correct wave behavior at the horizon and at infinity.
The parameters $N_x$ and $N_y$ specify the truncation of the expansions and the number of grid points chosen in the corresponding coordinate domains.
The double expansion then introduces a set of constants $C_{{\mathrm{I}},k,l}$ for each of the seven perturbation functions, where $k$ comes from the expansion in $x$ and $l$ from the expansion in $y$.

We choose the grid by employing Gauss-Lobato points for the radial coordinate $x$ 
\begin{eqnarray}
    x_I = \frac{1}{2} \left( 1 + \cos \left( \frac{I-1}{N_x -1} \pi \right) \right) \, , \quad I = 1, ..., N_x \, ,
\end{eqnarray}
and an equidistant mesh for the angular coordinate $y$
\begin{eqnarray}
    y_K = 2 \left( \frac{K-1}{N_y - 1} \right) - 1 \, , \quad K = 1, ..., N_y \, .
\end{eqnarray}
To evaluate the perturbation functions $X_I$ at the grid points, we need to interpolate the background functions and their derivatives at these grid points. 
In addition, from the 11 field equations for the perturbation functions, we only need to choose 7, while the remaining 4 can be used to test the accuracy of the solutions.
This leads us finally to a system of $7 \times N_x \times N_y$ algebraic equations to determine the constants $C_{{\mathrm{I}},k,l}$.
These algebraic equations can be put in the following form,
\begin{eqnarray}
    \left( \mathcal{M}_0 + \mathcal{M}_1 \omega + \mathcal{M}_2 \omega^2 \right) \Vec{C} = 0 \, ,
    \label{matrix_eq}
\end{eqnarray}
where $\mathcal{M}_0$, $\mathcal{M}_1$ and $\mathcal{M}_2$ are numerical matrices of size $(7 \times N_x \times N_y) \times (7 \times N_x \times N_y)$, $\vec{C}$ is the vector of constants $C_{{\mathrm{I}},k,l}$, and $\omega$ is the complex eigenvalue of a quasinormal mode.
We solve this quadratic eigenvalue problem (\ref{matrix_eq}) using Maple and Matlab with the Multiprecision Computing Toolbox Advanpix \cite{Advanpix}.

There are several ways to estimate the accuracy of the quasinormal modes calculated. For example, since from the 11 field equations only 7 need to be used to obtain the solutions, the remaining 4 are used to estimate the accuracy. In addition, other auxiliary parameters of the spectral decomposition can be varied to estimate the error (for example, the original number of points of the numerical background, the collocation grid, etc). We make this analysis for each of the modes calculated, and for all the results reported in this work. The relative precision of the quasinormal modes is always better than $10^{-5}$.
See \cite{Blazquez-Salcedo:2023hwg, Blazquez-Salcedo:2024oek, 
Blazquez-Salcedo:2024dur,
Berti:2025hly} for extensive details of the method.

\section{Spectrum of quasinormal modes for rotating shift-symmetric EsGB black holes}

Here we present the results for the quasinormal modes of spinning black holes in EsGB theory with shift symmetry. 
We focus on the fundamental modes of $(l=2,3)$-led polar, axial and scalar modes for $M_z=2$.
The characterization of the modes in terms of their leading multipolar $l$ behavior is based on an inspection of the perturbation functions, and also a smooth connection with the Kerr limit of the modes, as explained in \cite{Blazquez-Salcedo:2023hwg}.

\begin{figure*}[p!]
\begin{center}
\mbox{ 
\includegraphics[width=1.35\textwidth,angle=-90]{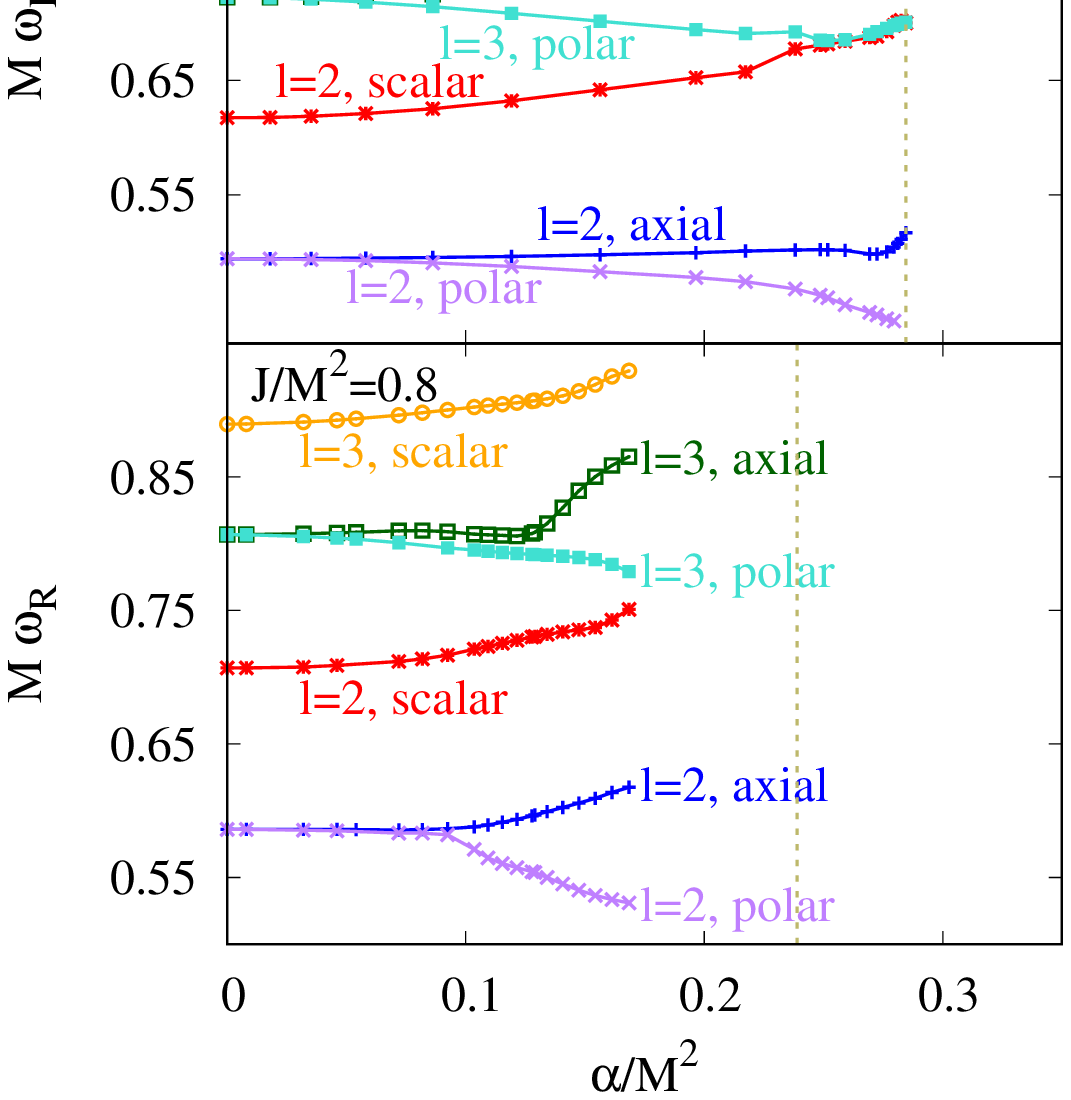}
\includegraphics[width=1.35\textwidth,angle=-90]{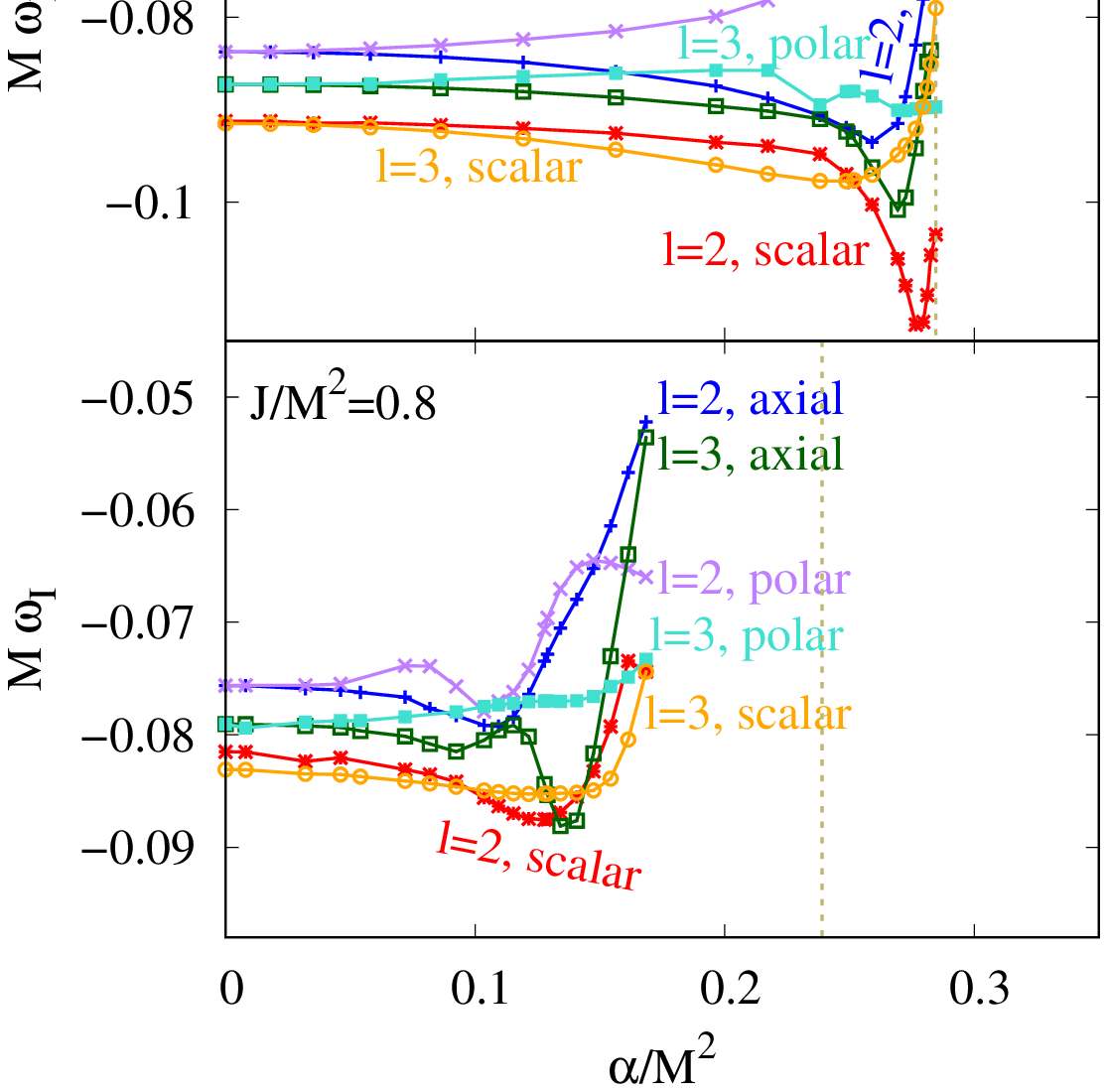}
}
\vspace*{-0.5cm}
\end{center}
\caption{
Fundamental shift-symmetric EsGB ($l=2$) and ($l=3$)-led modes for $M_z=2$: scaled real part $M\omega_R$ (left column) and scaled imaginary part $M\omega_I$ (right column) of the polar-led, axial-led, and scalar-led quasinormal modes as a function of the scaled coupling constant $\alpha/M^2$ for scaled angular momenta $J/M^2=0, 0.2, 0.4, 0.6, 0.8$ (from top to bottom).
The limiting coupling constant for each value of $J/M^2$ of the background black holes is shown by the dark khaki vertical line.
}
\label{fig_allj}
\end{figure*}

In figure \ref{fig_allj} we show the real and imaginary parts of the modes scaled by mass, respectively, $M\omega_R$ and $M\omega_I$, as a function of the scaled coupling constant $\xi=\alpha/M^2$.
The modes are computed for a range of scaled angular momenta $j=J/M^2=0$, 0.2, 0.4, 0.6, 0.8 which includes the static modes for this theory.
The dark khaki vertical lines in the figures indicate the limiting scaled coupling constant for each $j$, which decreases as $j$ grows.
These limiting values are derived from the domain of existence of the shift-symmetric EsGB black holes.
In the Appendix we provide the corresponding fitting functions for the modes in tables \ref{tab_fit1} - \ref{tab_fit_ult}.

Isospectrality of the modes is broken as the axial and polar modes no longer coincide like they do in GR, due to the presence of the non-trivial scalar field.
The $l=2$ polar modes give the lowest scaled real frequency irrespective of the scaled angular momentum and scaled coupling constant.
While the highest scaled real frequency is given by the $l=3$ scalar modes in all the cases considered.

In contrast, for the imaginary parts of the modes which 
determine the damping time of the modes via $\tau=1/|\omega_I|$,
the leading longest lived modes change drastically  
with the scaled angular momentum,
especially close to the limiting scaled coupling constant.

\begin{figure*}[p!]
\begin{center}
\mbox{ 
\includegraphics[width=1.35\textwidth,angle=-90]{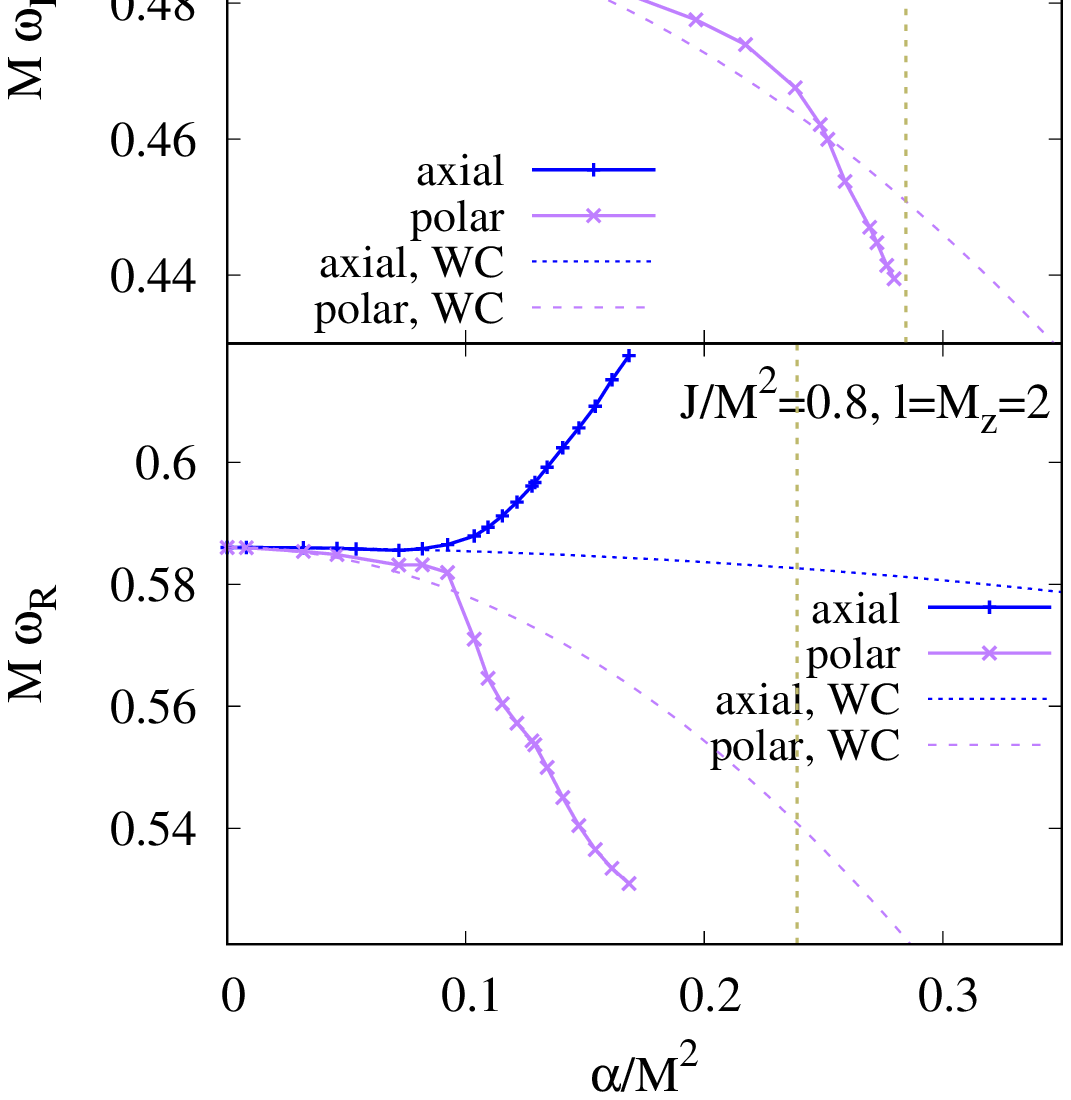}
\includegraphics[width=1.35\textwidth,angle=-90]{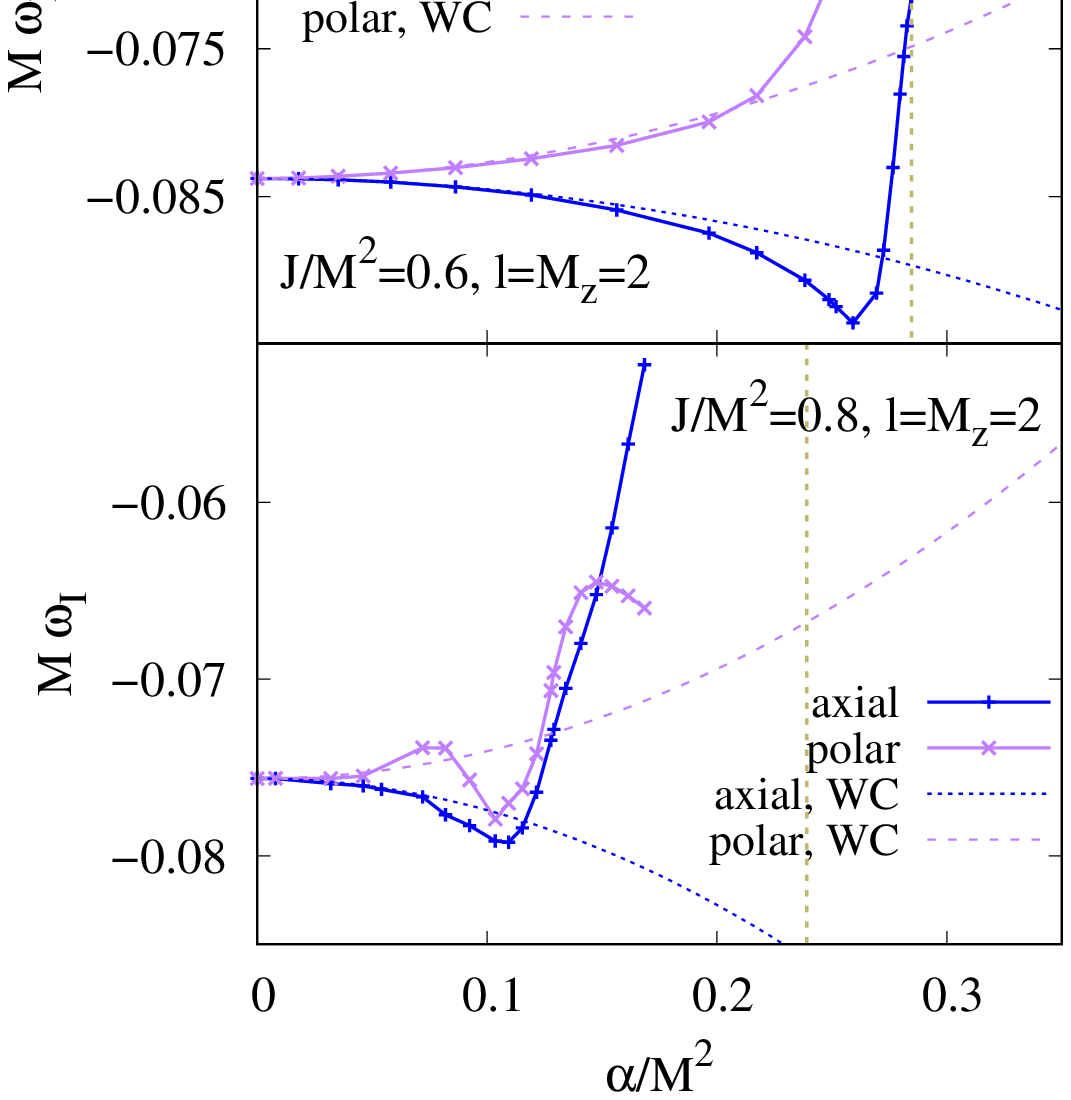}
}
\vspace*{-0.5cm}
\end{center}
\caption{
Comparison of the exact modes with $l=M_z=2$ fundamental polar-led and axial-led quasinormal modes where a weak coupling (WC) approximation was employed \cite{Chung:2024ira},
for ${J/M^2}=0.2,0.4,0.6,0.8$.
Also shown is a comparison between the static modes with a full coupling and the perturbative modes for a small scaled angular momentum ${J/M^2}=0.005$ (topmost) \cite{Chung:2024ira}.
}
\label{fig_allj_sc}
\end{figure*}

We show in figure \ref{fig_allj_sc} the comparison of our exact results with previous results obtained in a weak coupling (WC) approximation in \cite{Chung:2024vaf} for the $l=2$ gravitational modes.
The best perturbative result is seen for the scaled real part of the axial modes for $j=0.2$,
which agrees with ours up to about 85\% of the limiting scaled coupling constant.
The real part of the axial modes for $j=0.4$ and $j=0.6$ depicted in the figure is also in good agreement with ours for a wide range of the scaled coupling constant.
Besides, as shown in the topmost plot in figure \ref{fig_allj_sc}, the real part
of the perturbative axial modes for $j=0.005$ \cite{Chung:2024vaf} follows closely the exact static modes within a range similar to the $j=0.2$ case.

However, the results of the perturbative 
WC
calculations \cite{Chung:2024vaf} typically deviate from our exact calculations for larger values of the scaled coupling constant.
The dependence of the modes on the coupling constant then varies wildly especially in the imaginary part of the modes.
For instance, for a scaled black hole spin of $j=0.2$ we observe in figure \ref{fig_allj_sc}, that the largest deviation occurs around $\xi = 0.316$, a value that is close to the maximal value of the limiting scaled coupling constant. 
In particular, there are deviations in the imaginary part of the axial modes of about 5\%
and deviations in the real part of the polar modes of about 4\%.

As the black hole rotates faster, the corresponding limiting scaled coupling constant (indicated by the vertical line in figure \ref{fig_allj}) decreases.
Of all the cases we have considered, the largest deviation between our results and the 
WC
ones is 
found in figure \ref{fig_allj_sc} in the scaled imaginary part of the axial modes for $j=0.8$ at about $\xi=0.168$, which amounts to a deviation of about 55\%. 
This shows  
that the effect of the  
exact treatment of the coupling constant becomes 
highly relevant for rapidly rotating black holes.
On the other hand, the scaled real part of the perturbative axial modes agrees well with our results up to about $\xi=0.071$ (i.e., 30\% of the limiting scaled coupling constant). 
We summarize in table \ref{tab:threshold_coupling} the threshold of the coupling constant beyond which the perturbative computations \cite{Chung:2024ira, Chung:2024vaf} depart from our exact results. 
Due to a deterioration of our numerical precision,  
we exhibit the modes for $j=0.8$ only up to $\xi=0.168$, that is about 70\% of the limiting scaled coupling constant. 
The decline in the precision of the modes is mainly due to the decrease in the numerical precision of the background solutions, that deteriorates as we get closer and closer to the limit value of the coupling constant.
In particular, for this case, the estimated mean error of the background solutions is less than $10^{-5}$.
More precisely, it is around 
$10^{-8}$ close to GR, and around
$10^{-6}$ close to the limiting coupling constant.
Discussions about the limiting behavior of the solutions
can be found in \cite{Delgado:2020rev}.
As we discussed above, the precision of
all the modes obtained in this work is of $10^{-5}$ or better.

\begin{table}
    \centering
    \begin{tabular}{|c||c|c||c|c
    |}
 \hline
  \multirow{3}{*}{}  &
   \multicolumn{4}{|c|}{ Threshold of $\alpha/M^2$}  \\  
&  \multicolumn{2}{|c||}{ Axial} &
  \multicolumn{2}{|c|}{ Polar} 
\\ 
  $J/M^2$   & Real freq. & Im. freq. & Real freq. & Im. freq.
  \\
 \hline
$\leq$ 0.005	&	0.23	&	0.15	&	0.12	&	0.10		
\\
0.2	&	0.27	&	0.25	&	0.11	&	0.25	
\\
0.4	&	0.19	&	0.12	&	0.11	&	0.15	
\\
0.6	&	0.24	&	0.12	&	0.09	&	0.09	
\\
0.8	&	0.02	&	0.05	&	0.03	&	0.01	
\\
 \hline
    \end{tabular}
    \caption{
    Threshold of the scaled coupling constant $\alpha/M^2$ for each scaled angular momentum $J/M^2$ for the scaled real frequency $M\omega_R$ and imaginary frequency $M\omega_I$ of the $l=M_z=2$ fundamental axial-led and polar-led quasinormal modes, derived from the mode comparison in figure \ref{fig_allj_sc}. Beyond the threshold, results from the weak coupling approximation begin to deviate from exact results. Thresholds in the first row are obtained from a comparison between our exact static modes and the perturbative modes for $J/M^2=0.005$.}
    \label{tab:threshold_coupling}
\end{table}


\begin{figure*}[t!]
\begin{center}
\mbox{ 
\includegraphics[width=0.5\textwidth,angle=-90]{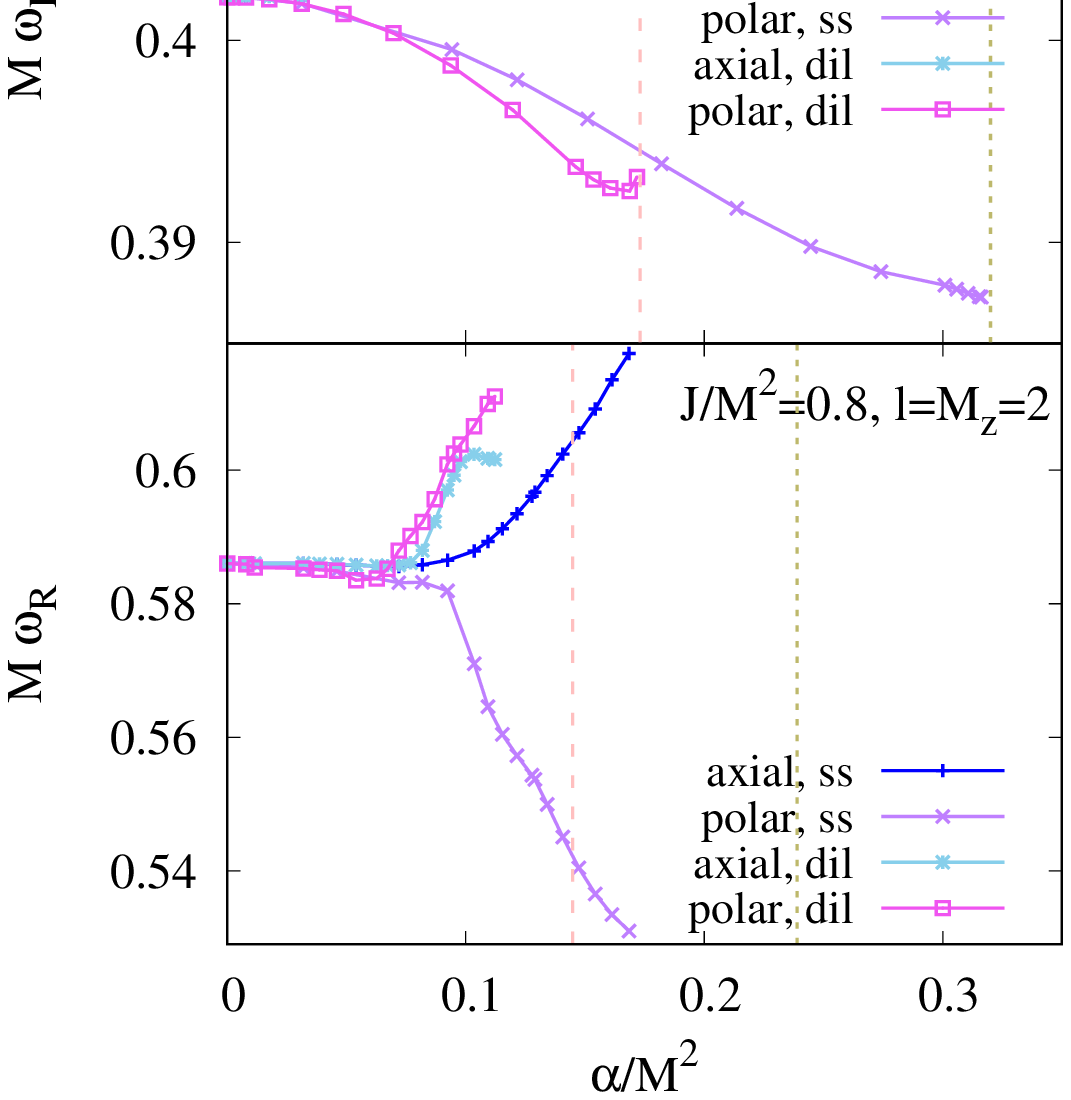}
\includegraphics[width=0.5\textwidth,angle=-90]{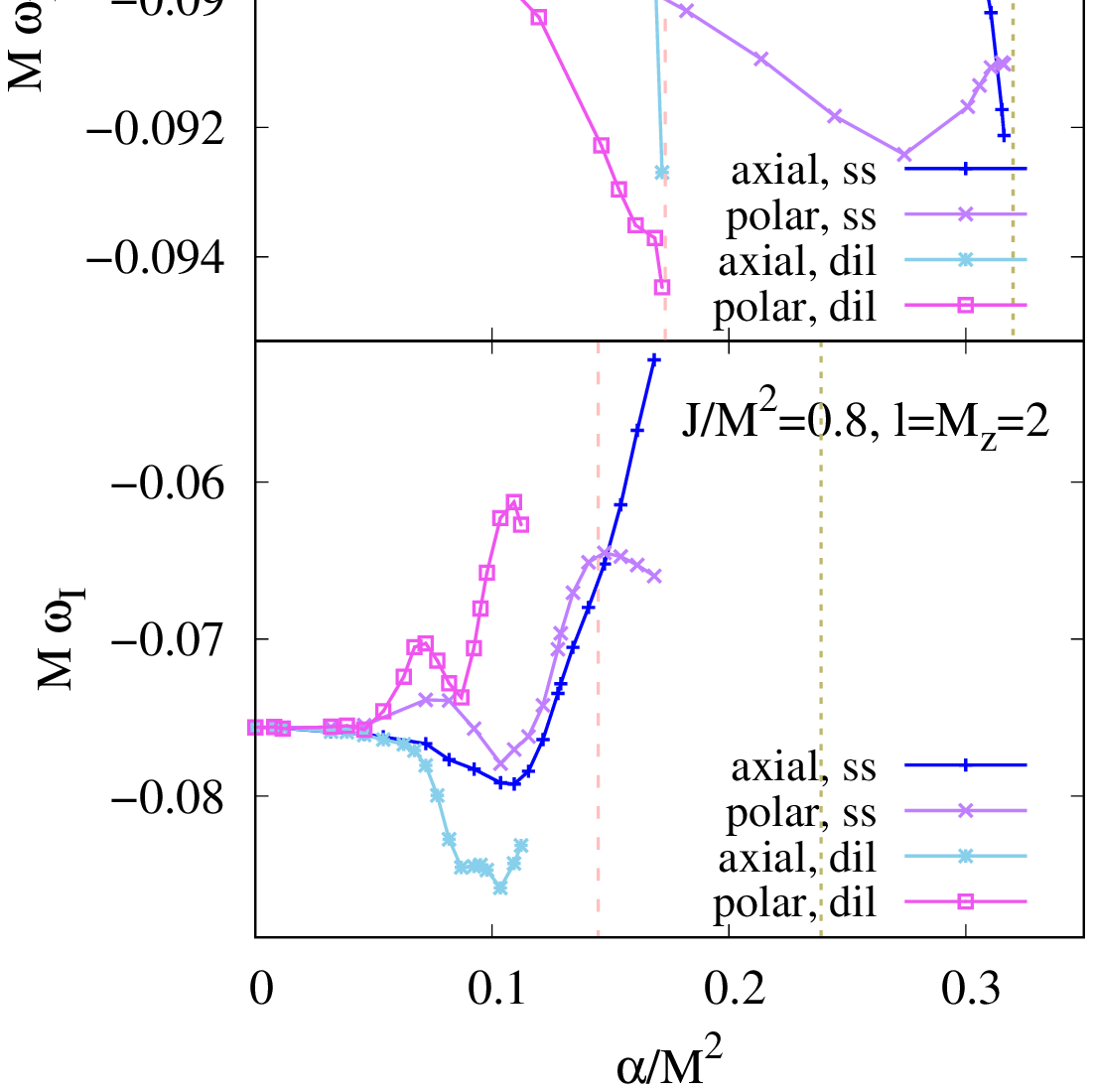}
}
\vspace*{-0.5cm}
\end{center}
\caption{
Comparison between the $l=M_z=2$ fundamental polar-led and axial-led modes of the shift-symmetric EsGB black holes (ss) and EdGB black holes (dil) \cite{Blazquez-Salcedo:2024oek, Blazquez-Salcedo:2024dur}, for ${J/M^2}=0.2, 0.8$.
}
\label{fig_ssdil_js}
\end{figure*} 

Lastly, let us compare the exact modes obtained here for the shift-symmetric EsGB theory
with those obtained for EdGB theory in \cite{Blazquez-Salcedo:2024oek, Blazquez-Salcedo:2024dur}.
Figure \ref{fig_ssdil_js} shows a comparison between the $l=M_z=2$ fundamental polar-led and axial-led modes for $j=0.2$ and $j=0.8$, respectively.
The pink vertical lines with a smaller value of the scaled coupling constant denote the limiting $\xi$ values for EdGB, while the dark khaki vertical lines denote those  
for the shift-symmetric EsGB.
We find that the modes of these two theories with a different coupling function agree well for
smaller coupling constants. 
This can be expected since the shift-symmetric EsGB has a linear coupling function, that corresponds to the first (relevant) term in the expansion of the EdGB coupling function.

In this work, we have examined only a fraction of the modes that we typically obtain using our method.  
By fixing $M_z$, the spectral method generally captures a spectrum of modes composed of higher $l$'s 
($l\ge M_z$) as well as higher excitation modes.
For instance, for the first excitation modes of the same $l$, their real parts are usually close to those of the fundamental modes, while their imaginary parts can be of an order of $10^{-1}$ larger in absolute value compared to the fundamental modes.

\section{Conclusions}

Following a series of work in the computation of quasinormal modes using our previously developed spectral method, we here obtained
the exact modes for rapidly rotating black holes in the shift-symmetric EsGB theory with a fully non-perturbative treatment of the coupling.  

We presented the results for the fundamental $l=2$ and $l=3$ modes for $M_z=2$, capturing all types of modes in the theory, namely the polar, axial and scalar-led modes, and covered the modes for almost the entire domain of the black hole solutions.
All the modes smoothly connect to the corresponding isospectral GR Kerr (Schwarzschild) modes for the rotating (static) EsGB black holes.
As the scaled coupling constant $\xi$ increases, the order of the modes changes, and thus the longest lived mode changes its character in a way that also depends
on the value of the scaled angular momentum.
Thus the dominant mode can be $l=2$ axial-led, polar-led or scalar-led, or even $(l=3)$-led.
However, these modes typically have quite large differences in the real part. Hence in principle it could be possible to determine the nature of the dominant mode by measuring the oscillation frequency, at least in some cases.
We note, that for large values of the scaled angular momentum such as $j=0.8$, the calculation of the modes becomes very sensitive towards the limiting $\xi$ value, limiting our numerical accuracy.
Therefore we showed the modes for this case only up to 70\% of the corresponding limiting value of $\xi$.

We compared our results with previous perturbative results obtained in a weak coupling approximation \cite{Chung:2024ira,Chung:2024vaf} for the $l=M_z=2$ gravitational modes.
Our results coincide well in the limit of small scaled coupling constant, whereas the disagreement grows strongly 
as we approach large values of the 
coupling. 
Finally we also compared 
our exact $l=M_z=2$ gravitational modes for the shift-symmetric EsGB and EdGB theories.
The domain of the background solutions and thus the modes in EdGB theory is smaller than that in the shift-symmetric EsGB theory.
As expected, the modes agree well, when $\xi$ is small, but 
behave rather differently towards the larger couplings as the non-linearity of the dilatonic coupling kicks in.

\section*{Acknowledgement}

We gratefully acknowledge support by DFG project Ku612/18-1, 
FCT project PTDC/FIS-AST/3041/2020, 
and MICINN project PID2021-125617NB-I00 ``QuasiMode''. JLBS gratefully acknowledges support from MICINN project CNS2023-144089 ``Quasinormal modes''.
FSK gratefully acknowledges support from ``Atracci\'on de Talento Investigador Cesar Nombela'' of the Comunidad de Madrid, grant no. 2024-T1/COM-31385.

\section{Appendix}

In this section we present the fitting functions for the quasinormal modes of black holes computed for $J/M^2=0, 0.2, 0.4, 0.6, 0.8$ in the shift-symmetric EsGB theory.
We find that a fit with only a constant 
and a quadratic term as in the weak coupling case \cite{Chung:2024ira,Chung:2024vaf} is no longer sufficient. 
We consider in general the following polynomial to fit the scaled real $M\omega_R$ and imaginary $M\omega_I$ part of the fundamental modes as a function of the scaled coupling constant $\xi$,
\begin{eqnarray}
    M\omega_{R,I} = \text{const.}  
    +   C_{\xi}\xi +
    C_{\xi^2}\xi^2    
    +   C_{\xi^3}\xi^3
    +   C_{\xi^4}\xi^4 +
    C_{\xi^5}\xi^5 \, .
    \label{poly_fit}
\end{eqnarray}
The first term of the polynomial is a constant which is essentially the Schwarzschild/Kerr mode,
and $C_{\xi^a}$'s are the corresponding coefficients 
for the scaled coupling constant
with power of $a=1,..,5$.
For most of the cases, the corresponding fitting functions do not require all the terms in (\ref{poly_fit}) for best fitting, and in fact only very few cases involve the fifth order term in (\ref{poly_fit}).

As rotation increases,
and as the coupling strength grows (or as we approach the domain limit),
obtaining an analytical fit of the data
gets more challenging. 
This is because of the dependence of the mode on the coupling constant that becomes volatile. 
Therefore we provide in the tables \ref{tab_fit1} - \ref{tab_fit_ult} below also the range
of the $\xi$ up to where we 
obtain the best fit.
Note that some of the fits
were obtained up to the limit we reached with our quasinormal modes.
As seen from the following tables, for most of the cases, a fit higher than second order (with at times an inclusion of a linear term) applies for a larger range of $\xi$, compared to the one in the weak coupling approximation (see table \ref{tab:threshold_coupling}).


\begin{table}[h!]
\begin{center}
\begin{tabular}{ | c || c | c| c| c  || c |}
 \hline
 & const.  & $C_{\xi^2}$ & $C_{\xi^3}$ & $C_{\xi^4}$  & up to $\xi$ 
\\ 
  \hline
$l=2$, axial & 0.37367 
& 0.068387
& -0.055620
& 
& 0.29
  \\ 
   \hline
  $l=2$, polar & 0.37367 
& -0.23993
& 0.21673
& 0.57363
& 0.29
\\ 
   \hline
$l=3$, axial & 0.59944 
& 0.18955
& -0.82486
& 2.26121
& 0.32
  \\ 
  \hline
$l=3$, polar & 0.59944 
& -1.04125
& 3.29484
& -3.62983
& 0.32
  \\ 
  \hline
$l=2$, scalar & 0.48381 
& 0.66547
& 
& 1.08601
& 0.32
  \\ 
  \hline
$l=3$, scalar & 0.67529 
& 1.78553
& -3.48822
& 6.54248
& 0.32
  \\ 
  \hline
\end{tabular}
\end{center}
\caption{$J/M^2=0$: Fitting functions for the scaled real part $M\omega_R$ of the axial-led, polar-led, and scalar-led fundamental $(l=2)$ and $(l=3)$-led modes for $M_z=2$. } 
\label{tab_fit1}
\end{table}



\begin{table}[h!]
\begin{center}
\begin{tabular}{ | c || c| c| c| c| c| c || c |}
 \hline
 & const. & $C_{\xi}$ & $C_{\xi^2}$ & $C_{\xi^3}$ & $C_{\xi^4}$ & $C_{\xi^5}$ & up to $\xi$ 
\\ 
  \hline
$l=2$, axial & -0.088962 
&  0.00019122
& 
& 0.058767
& 
& -1.40519
& 0.15
  \\ 
   \hline
  $l=2$, polar & -0.088962 
&  -0.0022980
& 
& -0.61401
& 1.78256
&
& 0.29
\\ 
   \hline
$l=3$, axial & -0.092703 
&  -0.00045510
& 0.028085
& -0.30585
& 1.72615
& -3.62043
& 0.29
  \\ 
  \hline
$l=3$, polar & -0.092703 
&  -0.0010525
& -0.10107
& 0.26903
& 
& 
& 0.29
  \\ 
  \hline
$l=2$, scalar & -0.096769 
&  
& 0.094095
& -0.41625
& 1.46514
& 
& 0.32
  \\ 
  \hline
$l=3$, scalar & -0.096537 
&  
& 0.16185
& -1.11991
& 2.84158
& 
& 0.32
  \\ 
  \hline
\end{tabular}
\end{center}
\caption{
$J/M^2=0$: Fitting functions for the scaled imaginary part $M\omega_I$ of the axial-led, polar-led, and scalar-led fundamental $(l=2)$ and $(l=3)$-led modes for $M_z=2$.
} 
\end{table}



\begin{table}[h!]
\begin{center}
\begin{tabular}{ | c ||  c| c| c| c| c || c |}
 \hline
 & const. & $C_{\xi}$  & $C_{\xi^2}$ & $C_{\xi^3}$ & $C_{\xi^4}$  & up to $\xi$ 
\\ 
  \hline
$l=2$, axial & 0.40215 
&  0.0040144
& 
& 0.63326
& -1.33504
& 0.27
  \\ 
   \hline
  $l=2$, polar & 0.40215 
&  
& -0.30116
& 
& 1.58927
& 0.30
\\ 
   \hline
$l=3$, axial & 0.62968 
&  
& 0.22664
& -0.88789
& 2.81144
& 0.31
  \\ 
  \hline
$l=3$, polar & 0.62968 
&  
& -1.13181
& 3.17165
& -2.50049
& 0.31
  \\ 
  \hline
$l=2$, scalar & 0.51712 
& 0.011722 
& 0.72068
& 
& 1.51998
& 0.31
  \\ 
  \hline
$l=3$, scalar & 0.70858 
&  
& 1.59011
& 
& -2.83220
& 0.32
  \\ 
  \hline
\end{tabular}
\end{center}
\caption{
$J/M^2=0.2$: Fitting functions for the scaled real part $M\omega_R$ of the axial-led, polar-led, and scalar-led fundamental $(l=2)$ and $(l=3)$-led modes for $M_z=2$.
} 
\end{table}



\begin{table}[h!]
\begin{center}
\begin{tabular}{ | c || c| c| c| c| c || c |}
 \hline
 & const. & $C_{\xi}$ & $C_{\xi^2}$ & $C_{\xi^3}$ & $C_{\xi^4}$  & up to $\xi$ 
\\ 
  \hline
$l=2$, axial & -0.088311 
&  0.00059332
& 
& 0.12699
& -0.29897
& 0.25
  \\ 
   \hline
  $l=2$, polar & -0.08831 
&  -0.001807
& 
& -0.4325
& 0.932
& 0.27
\\ 
   \hline
$l=3$, axial & -0.092124 
&  
& 0.016447
& -0.079903
& 0.23416
& 0.27
  \\ 
  \hline
$l=3$, polar & -0.092124 
&  -0.0019111
& -0.054688
& 
& 0.49815
& 0.27
  \\ 
  \hline
$l=2$, scalar & -0.096382 
&  
& 0.077734
& -0.43800
& 1.47117
& 0.27
  \\ 
  \hline
$l=3$, scalar & -0.096151 
&  
& 0.071679
& -0.45123
& 1.00643
& 0.25
  \\ 
  \hline
\end{tabular}
\end{center}
\caption{
$J/M^2=0.2$: Fitting functions for the scaled imaginary part $M\omega_I$ of the axial-led, polar-led, and scalar-led fundamental $(l=2)$ and $(l=3)$-led modes for $M_z=2$.
} 
\end{table}



\begin{table}[h!]
\begin{center}
\begin{tabular}{ | c || c| c| c| c| c || c |}
 \hline
 & const.  & $C_{\xi^2}$ & $C_{\xi^3}$ & $C_{\xi^4}$ & $C_{\xi^5}$ & up to $\xi$ 
\\ 
  \hline
$l=2$, axial & 0.43984 
& 0.12943
& 0.057970
& 
& 
& 0.30
  \\ 
   \hline
  $l=2$, polar & 0.43984 
& -0.43093
& 
& 4.75327
&-10.3132
& 0.28
\\ 
   \hline
$l=3$, axial & 0.66892 
& 0.34128
& -1.751
& 6.94337
& 
& 0.30
  \\ 
  \hline
$l=3$, polar & 0.66892 
& -1.35354
& 4.30287
& -4.2662
& 
& 0.31
  \\ 
  \hline
$l=2$, scalar & 0.55950 
& 1.0092
& 
& -1.28804
& 
& 0.31
  \\ 
  \hline
$l=3$, scalar & 0.75052
& 2.14602
& -3.21421
& 
& 
& 0.30
  \\ 
  \hline
\end{tabular}
\end{center}
\caption{
$J/M^2=0.4$: Fitting functions for the scaled real part $M\omega_R$ of the axial-led, polar-led, and scalar-led fundamental $(l=2)$ and $(l=3)$-led modes for $M_z=2$.
} 
\end{table}



\begin{table}[h!]
\begin{center}
\begin{tabular}{ | c || c| c| c| c| c || c |}
 \hline
 & const. & $C_{\xi}$ & $C_{\xi^2}$ & $C_{\xi^3}$ & $C_{\xi^4}$  & up to $\xi$ 
\\ 
  \hline
$l=2$, axial & -0.086882 
&  
& -0.018920
& 0.19278
& -0.78610
& 0.28
  \\ 
   \hline
  $l=2$, polar & -0.086859 
&  
& 
& 0.21001
& -1.40788
& 0.30
\\ 
   \hline
$l=3$, axial & -0.090620 
&  
& 
& -0.072521
& 0.22987
& 0.21
  \\ 
  \hline
$l=3$, polar & -0.090620 
&  -0.0012312
& -0.0065374
& 
& 
& 0.21
  \\ 
  \hline
$l=2$, scalar & -0.094913 
&  
& -0.0076154
& 0.11121
& 
& 0.30
  \\ 
  \hline
$l=3$, scalar & -0.094812 
&  
& -0.015924
& 
& -0.44815
& 0.30
  \\ 
  \hline
\end{tabular}
\end{center}
\caption{
$J/M^2=0.4$: Fitting functions for the scaled imaginary part $M\omega_I$ of the axial-led, polar-led, and scalar-led fundamental $(l=2)$ and $(l=3)$-led modes for $M_z=2$.
} 
\end{table}



\begin{table}[h!]
\begin{center}
\begin{tabular}{ | c || c|  c| c| c || c |}
 \hline
 & const.  & $C_{\xi^2}$ & $C_{\xi^3}$ & $C_{\xi^4}$  & up to $\xi$ 
\\ 
  \hline
$l=2$, axial & 0.49406 
& 0.13728
& 
& 
& 0.24
  \\ 
   \hline
  $l=2$, polar & 0.49404 
& -0.80871
& 4.14472
& -11.2605
& 0.24
\\ 
   \hline
$l=3$, axial & 0.72280 
& 0.91093
& -8.34835
& 31.4192
& 0.26
  \\ 
  \hline
$l=3$, polar & 0.72288 
& -1.55968
& 5.65631
& -7.32024
& 0.25
  \\ 
  \hline
$l=2$, scalar & 0.61736 
& 1.14805
& 
& -6.26948
& 0.22
  \\ 
  \hline
$l=3$, scalar & 0.80644 
& 2.3968
& -7.49936
& 10.2467
& 0.25
  \\ 
  \hline
\end{tabular}
\end{center}
\caption{
$J/M^2=0.6$: Fitting functions for the scaled real part $M\omega_R$ of the axial-led, polar-led, and scalar-led fundamental $(l=2)$ and $(l=3)$-led modes for $M_z=2$.
} 
\end{table}



\begin{table}[h!]
\begin{center}
\begin{tabular}{ | c || c| c| c| c| c || c |}
 \hline
 & const. & $C_{\xi}$ & $C_{\xi^2}$ & $C_{\xi^3}$ & $C_{\xi^4}$  & up to $\xi$ 
\\ 
  \hline
$l=2$, axial & -0.083765 
&  
& -0.15513
& 1.13554
& -4.21119
& 0.26
  \\ 
   \hline
  $l=2$, polar & -0.083765 
&  
& 0.13921
& -0.67965
& 2.41021
& 0.20
\\ 
   \hline
$l=3$, axial & -0.087280 
&  
& -0.049691
& 
& -0.27259
& 0.24
  \\ 
  \hline
$l=3$, polar & -0.087280 
&  
& 0.069483
& 
& -0.78354
& 0.22
  \\ 
  \hline
$l=2$, scalar & -0.091245 
&  
& -0.049609
& 
& -0.21400
& 0.24
  \\ 
  \hline
$l=3$, scalar & -0.091501 
&  -0.0040743
& 
& -0.94078
& 2.31049
& 0.24
  \\ 
  \hline
\end{tabular}
\end{center}
\caption{
$J/M^2=0.6$: Fitting functions for the scaled imaginary part $M\omega_I$ of the axial-led, polar-led, and scalar-led fundamental $(l=2)$ and $(l=3)$-led modes for $M_z=2$.
} 
\end{table}



\begin{table}[h!]
\begin{center}
\begin{tabular}{ | c || c| c| c| c| c| c || c |}
 \hline
 & const. & $C_{\xi}$ & $C_{\xi^2}$ & $C_{\xi^3}$ & $C_{\xi^4}$ & $C_{\xi^5}$ & up to $\xi$ 
\\ 
  \hline
$l=2$, axial & 0.58602 
&  0.0095338
& 
& -12.4504
& 130.262
& 
& 0.13
  \\ 
   \hline
  $l=2$, polar & 0.58602 
&  
& -1.23231
& 23.1094
& -161.605
&
& 0.09
\\ 
   \hline
$l=3$, axial & 0.80678 
&  
& 
& 23.5558
& -224.515
& 
& 0.10
  \\ 
  \hline
$l=3$, polar & 0.80678 
&  
& 
& -40.7329
& 413.653
& -1177.98
& 0.15
  \\ 
  \hline
$l=2$, scalar & 0.70682 
&  
& 1.7076
& -44.4424
& 644.965
& -2462.09
& 0.14
  \\ 
  \hline
$l=3$, scalar & 0.88948 
&  
& 
& 53.7799
& -642.354
& 2240.56
& 0.16
  \\ 
  \hline
\end{tabular}
\end{center}
\caption{
$J/M^2=0.8$: Fitting functions for the scaled real part $M\omega_R$ of the axial-led, polar-led, and scalar-led fundamental $(l=2)$ and $(l=3)$-led modes for $M_z=2$.
} 
\end{table}



\begin{table}[h!]
\begin{center}
\begin{tabular}{ | c || c| c| c| c| c| c || c |}
 \hline
 & const. & $C_{\xi}$ & $C_{\xi^2}$ & $C_{\xi^3}$ & $C_{\xi^4}$ & $C_{\xi^5}$ & up to $\xi$ 
\\ 
  \hline
$l=2$, axial & -0.075630 
&  
& 
& -5.41865
& 21.6672
& 
& 0.10
  \\ 
   \hline
  $l=2$, polar & -0.075630 
&  0.0043261
& -0.38365
& 8.40446
& -15.0295
&
& 0.08
\\ 
   \hline
$l=3$, axial & -0.079046 
&  
& -0.13169
& 
& -18.2899
& 
& 0.09
  \\ 
  \hline
$l=3$, polar & -0.079046 
&  
& 0.35631
& -9.84932
& 126.068
& -493.866
& 0.13
  \\ 
  \hline
$l=2$, scalar & -0.081520 
&  
& -2.09294
& 67.95
& -801.508
& 2939.41
& 0.15
  \\ 
  \hline
$l=3$, scalar & -0.083091 
&  
& -0.21802
& 
& 
& 41.0615
& 0.15
  \\ 
  \hline
\end{tabular}
\end{center}
\caption{
$J/M^2=0.8$: Fitting functions for the scaled imaginary part $M\omega_I$ of the axial-led, polar-led, and scalar-led fundamental $(l=2)$ and $(l=3)$-led modes for $M_z=2$.
} 
\label{tab_fit_ult}
\end{table}

\end{document}